# Biaxial Gaussian Beams, Hermite-Gaussian Beams, and Laguerre-Gaussian Vortex Beams in Isotropy-Broken Materials.

Maxim Durach (mdurach@georgiasouthern.edu)

Center for Advanced Materials Science, Department of Biochemistry, Chemistry & Physics, Georgia Southern University, Statesboro, GA 30460, USA

**Abstract.** We develop the paraxial approximation for electromagnetic fields in arbitrary isotropy-broken media in terms of the ray-wave tilt and the curvature of materials' Fresnel wave surfaces. We obtain solutions of the paraxial equation in the form of biaxial Gaussian beams, which is a novel class of electromagnetic field distributions in generic isotropy-broken materials. Such beams have been previously observed experimentally and numerically in hyperbolic metamaterials but evaded theoretical analysis in the literature up to now. The biaxial Gaussian beams have two axes: one in the direction of Abraham momentum, corresponding to the ray propagation, and another in the direction of Minkowski momentum, corresponding to the wave propagation, in agreement with the recent theory of refraction, ray-wave tilt, and hidden momentum [Durach, 2024, Ref. 1]. We show that the curvature of the wavefronts in the biaxial Gaussian beams correspond to the curvature of the Fresnel wave surface at the central wave vector of the beam. We obtain the higher-order modes of the biaxial beams, including the biaxial Hermite-Gaussian and Laguerre-Gaussian vortex beams, which opens avenues toward studies of optical angular momentum (OAM) in isotropy-broken media, including generic anisotropic and bianisotropic materials.

*1. Introduction*

In this paper, which is an extension of Ref. [1], we continue to conceptualize and describe isotropy-broken materials as a broad class of electromagnetic media, which do not feature isotropy. They are characterized by Fresnel wave surfaces without spherical symmetry and by non-transverse electromagnetic fields. They include all anisotropic and bianisotropic metamaterials [2-7].

It is well known from works of Fermat and Huygens in the XVII century, that rays are directed perpendicular to the wavefronts in isotropic media. Correspondingly, when considering a paraxial beam of rays, the wavefronts propagate close-to-parallel with the beam axis. The non-parallel to beam propagation of wavefronts in uniaxial media has been discussed [8,9] and numerically visualized in beams refracted into hyperbolic metamaterials [10,11], but an analytical theory of this effect is missing from the literature. This effect corresponds to a general phenomenon of ray-wave tilt due to non-parallel Abraham and Minkowski momenta and presence of tangential component of hidden momentum and bound charge waves in isotropy-broken materials [1]. In this manuscript we obtain the paraxial equation for isotropy-broken media and express its solutions as biaxial Gaussian beams, with one axis corresponding to the direction of the ray and another to wavefront propagation.

The Fresnel wave surfaces of isotropic media are spherically symmetric with negative curvatures inversely proportional to the wavelength $\lambda$ [12]. Correspondingly, Gaussian beams in isotropic media are expected to have converging wavefronts before foci and diverging wavefronts after, which is the case in all conventional media with positive indices of refraction. Surprisingly, it was demonstrated that negative-index isotropic media feature converging wavefronts after passing the focal plane [13]. This effect has also been numerically observed in the emission of sources into hyperbolic metamaterials [14,15], but also evaded theoretical analysis in the literature. In this manuscript we directly relate the curvature of the Fresnel wave surface to the curvature of the wavefronts in the biaxial Gaussian beams. For example, open-topology Fresnel wave surfaces of multihypebolic media [16-25] feature positive curvature which results in Gaussian beams with converging wavefronts after foci.

Orbital angular momentum (OAM) of light in isotropic media has occupied central role in photonics recently with such potential applications as communications, sensing, and optical manipulation [26,27]. Nevertheless, there are no studies of OAM of electromagnetic fields inside isotropy-broken materials. We obtain high-order modes in the form of biaxial Hermite-Gaussian and Laguerre-Gaussian vortex beams with topological charge. This result opens a new frontier for investigations of OAM in isotropy-broken materials.

## 2. Paraxial Equation and biaxial Gaussian beams in isotropy-broken materials.

Consider a plane wave with wave vector $\boldsymbol{k} = (k_x, k_y, k_z)$ and frequency $\omega = k_0 c$ propagating through an isotropy-broken material described by bianisotropic constitutive relations:

$$\begin{pmatrix} \boldsymbol{D} \\ \boldsymbol{B} \end{pmatrix} = \widehat{M} \begin{pmatrix} \boldsymbol{E} \\ \boldsymbol{H} \end{pmatrix} = \begin{pmatrix} \hat{\epsilon} & \hat{X} \\ \hat{Y} & \hat{\mu} \end{pmatrix} \begin{pmatrix} \boldsymbol{E} \\ \boldsymbol{H} \end{pmatrix}, \tag{1}$$

The properties of such wave are described using the characteristic matrix method. The z-component of the wave vector and the transverse field satisfy the following eigenproblem

$$\widehat{\Delta}(k_x, k_y) \cdot (E_x, E_y, H_x, H_y)^T = \left(\frac{k_z}{k_0}\right) (E_x, E_y, H_x, H_y)^T \tag{2}$$

The characteristic matrix of an isotropy-broken material given by Eq. (1) is [23,25]

$$\widehat{\Delta}(k_x, k_y) = \widehat{\Delta} = \widehat{P}^{-1} \cdot \left(\widehat{M}_{\parallel,\parallel} - (\widehat{M}_{\parallel,z} + \hat{q}^T) \cdot \widehat{M}_{z,z}^{-1} \cdot (\widehat{M}_{z,\parallel} + \hat{q})\right) \tag{3}$$

where additional matrices were defined in terms of vector $\boldsymbol{q} = \boldsymbol{k}/k_0$

$$\widehat{M}_{\parallel,\parallel} = \begin{pmatrix} \epsilon_{11} & \epsilon_{12} & X_{11} & X_{12} \\ \epsilon_{21} & \epsilon_{22} & X_{21} & X_{22} \\ Y_{11} & Y_{12} & \mu_{11} & \mu_{12} \\ Y_{21} & Y_{22} & \mu_{21} & \mu_{22} \end{pmatrix}, \widehat{M}_{\parallel,z} = \begin{pmatrix} \epsilon_{13} & X_{13} \\ \epsilon_{23} & X_{23} \\ Y_{13} & \mu_{13} \\ Y_{23} & \mu_{23} \end{pmatrix},$$

$$\widehat{P} = \widehat{P}^{-1} = \begin{pmatrix} 0 & 0 & 0 & -1 \\ 0 & 0 & 1 & 0 \\ 0 & 1 & 0 & 0 \\ -1 & 0 & 0 & 0 \end{pmatrix}, \hat{q} = \begin{pmatrix} 0 & 0 & -q_y & q_x \\ q_y & -q_x & 0 & 0 \end{pmatrix}$$

The longitudinal fields satisfy

-2-

$$(E_z, H_z)^T = -\widehat{M}_z^{-1} \cdot (\widehat{M}_{z\|} + \hat{q}) \cdot (E_x, E_y, H_x, H_y)^T \tag{4}$$

The characteristic equation for the eigenvalue problem of Eq. (2) is a quartic equation

$$\mathcal{H}(\boldsymbol{k}, k_0) = q_z^4 - \text{tr}\left(\widehat{\Delta}(\boldsymbol{q}_\perp)\right) q_z^3 - \xi(\boldsymbol{q}_\perp) q_z^2 - \zeta(\boldsymbol{q}_\perp) q_z + \det\left(\widehat{\Delta}(\boldsymbol{q}_\perp)\right) = 0 \tag{5}$$

where $\xi = \frac{1}{2}\left(\text{tr}(\widehat{\Delta}^2) - \text{tr}(\widehat{\Delta})^2\right)$, $\zeta = \frac{1}{6}\left(2\,\text{tr}(\widehat{\Delta}^3) - 3\,\text{tr}(\widehat{\Delta}^2)\,\text{tr}(\widehat{\Delta}) + \text{tr}(\widehat{\Delta})^3\right)$.

At given $\boldsymbol{k}_\perp = (k_x, k_y)$ Eq. (5) has 4 roots $k_z = k_z^{(i)}$, $i = 1,..4$ corresponding to the points $\left(k_x, k_y, k_z^{(i)}\right)$ at the Fresnel wave surface. For plane wave propagating paraxially to the z-axis

$$k_z^{(i)}(\boldsymbol{k}_\perp) = k_z^{(i)}(0) + \boldsymbol{k}_\perp^T \cdot \boldsymbol{t} + \frac{1}{2!} \boldsymbol{k}_\perp^T \cdot \widehat{H} \cdot \boldsymbol{k}_\perp + \cdots \tag{6}$$

where $\boldsymbol{t} = \left(\partial_{k_x}\left[k_z^{(i)}(\boldsymbol{k}_\perp)\right], \partial_{k_y}\left[k_z^{(i)}(\boldsymbol{k}_\perp)\right]\right)$ and $\widehat{H}_{ij} = \frac{\partial^2 k_z^{(i)}(\boldsymbol{k}_\perp)}{\partial k_i \partial k_j}$ are the gradient and Hessian matrix respectively. Similarly, the eigenvectors of Eq. (3) can be expressed as

$$\Gamma_\perp^{(i)}(\boldsymbol{k}_\perp) = \Gamma_\perp^{(i)}(\boldsymbol{0}) + \hat{J} \cdot \boldsymbol{k}_\perp + \cdots \tag{7}$$

where $\hat{J} = \left(\frac{\partial \Gamma_\perp^{(i)}}{\partial k_x}, \frac{\partial \Gamma_\perp^{(i)}}{\partial k_y}\right)$ is the Jacobian matrix.

We consider a Gaussian beam with a focus in $z = 0$ plane. Its transverse field $\Gamma_\perp^{(i)} = (E_x, E_y, H_x, H_y)^T$ can be expressed as

$$\Gamma_\perp^{(i)}(\boldsymbol{r}_\perp, z) = w_0 \int \frac{d\boldsymbol{k}_\perp}{(2\pi)^2} e^{i\boldsymbol{k}_\perp \boldsymbol{r}_\perp} e^{-\frac{k_\perp^2 w^2}{4}} \Gamma_\perp^{(i)}(\boldsymbol{k}_\perp) \cdot \exp\left(i k_z^{(i)}(\boldsymbol{k}_\perp) z\right) \tag{8}$$

Substituting the paraxial approximation of Eqs. (6)-(7) into Eq. (8) we obtain

$$\Gamma_\perp^{(i)}(\boldsymbol{r}_\perp, z) = \Gamma_\perp^{(i)}(\boldsymbol{0}) \cdot e^{i k_z^{(i)}(0) z} \cdot G(\boldsymbol{r}) - i \cdot e^{i k_z^{(i)}(0) z} \left(\hat{J} \cdot \boldsymbol{\nabla}_{\boldsymbol{r}_\perp}\right) \cdot G(\boldsymbol{r})$$

where the second term is considered small. The function $G(\boldsymbol{r})$ is given by

$$G(\boldsymbol{r}) = w_0 \int \frac{d\boldsymbol{k}_\perp}{(2\pi)^2} e^{i\boldsymbol{k}_\perp \boldsymbol{\rho}} \cdot \exp(\boldsymbol{k}_\perp^T \cdot \hat{\alpha}(z) \cdot \boldsymbol{k}_\perp), \quad \hat{\alpha}(z) = -\frac{w_0^2}{4} \hat{1} + \frac{i}{2!} \widehat{H} z \tag{9}$$

and $\boldsymbol{\rho}(z) = (x + t_x z, y + t_y z)$. Evaluating the integral in Eq. (9) we obtain the desired expression for the *biaxial Gaussian beam profile* in the arbitrary isotropy-broken material as

$$G(\boldsymbol{r}) = \frac{1}{4\pi} \frac{w_0}{\sqrt{\det(\hat{\alpha}(z))}} \exp\left(\frac{1}{4} \boldsymbol{\rho}(z) \cdot \hat{\alpha}(z)^{-1} \cdot \boldsymbol{\rho}^T(z)\right) \tag{10}$$

Note that function $G(\boldsymbol{r})$ satisfies the *generalized paraxial equation in isotropy-broken media*

$$i \frac{\partial G(\boldsymbol{r})}{\partial z} - \frac{1}{2!} \boldsymbol{\nabla}_\rho \cdot \left(\widehat{H} \boldsymbol{\nabla}_\rho G(\boldsymbol{r})\right) = 0 \tag{11}$$

The expressions Eq. (10)-(11) become especially transparent in the coordinate system where the Hessian matrix of the Fresnel wave surface $\hat{H}$ is diagonal $\hat{H} = r_x \hat{x}\hat{x} + r_y \hat{y}\hat{y}$, where coefficients $r$ characterize the curvatures of the Fresnel wave surface in the direction of the wavefronts propagation. In these notations Eq. (11) turns into

$$i \frac{\partial G(r)}{\partial z} - \frac{r_x}{2!}\frac{\partial^2 G(r)}{\partial^2 \rho_x} - \frac{r_y}{2!}\frac{\partial^2 G(r)}{\partial^2 \rho_y} = 0 \tag{12}$$

while the solution Eq. (10) is expressed as

$$G(\mathbf{r}) = g(x, t_x, r_x) g(y, t_y, r_y), \tag{13}$$

$$g(x, t_x, r_x) = \frac{w_0^{1/2}}{2\pi^{1/2}} \frac{1}{\sqrt{\left(\frac{w_0^2}{4} - \frac{i}{2!}r_x z\right)}} \exp\left(-\frac{1}{4}\frac{(x+t_x z)^2}{\frac{w_0^2}{4} - \frac{i}{2!}r_x z}\right) \tag{14}$$

We can rewrite Eq. (14) as

$$g(x, t_x, r_x) = \frac{1}{\pi^{1/2}} \frac{1}{\sqrt{w_x(z)}} \exp\left(-\frac{(x+t_x z)^2}{w_x^2(z)}\right) \exp\left(i\left\{\frac{k_z(x+t_x z)^2}{2R_x(z)} - \frac{1}{2}\psi_{G_x}\right\}\right), \tag{15}$$

where Gouy phase, beam radius, and wavefront radius of curvature in x-direction are respectively

$$\psi_{G_x} = \arctan\left(\frac{z}{z_x}\right), \quad w_x(z) = w_0\sqrt{1+\left(\frac{z}{z_x}\right)^2}, \quad R_x(z) = -r_x k_z z \left(1+\left(\frac{z_x}{z}\right)^2\right). \tag{16}$$

The Rayleigh length in the x-direction is $z_x = -\frac{w_0^2}{2r_x}$.

As we demonstrate in Fig. 1 and as can be seen from Eqs. (16) all the parameters of the beam depend on the curvature of the Fresnel wave surface. Specifically, the Rayleigh length $z_x$ and correspondingly the confocal parameter $2z_x$ are inversely proportional to the curvature of the Fresnel wave surface. Fresnel wave surfaces of isotropic media have spherical symmetry and their curvatures are negative $r_x < 0$. In this case the wavefront radius of curvature is positive $R_x > 0$, as can be seen in Fig. 1(a). If the curvature of the Fresnel wave surface is positive $r_x > 0$, the wavefront radius of curvature is negative $R_x < 0$, corresponding to converging wavefronts after passing the focal plane as shown in Fig. 1(b)-(c). Converging wavefronts in Gaussian beams have been previously discussed in relation to isotropic negative index materials [13], but never for isotropy-broken media. Please note that if $r_x > 0$, the sign of the accumulated Gouy phase is negative.

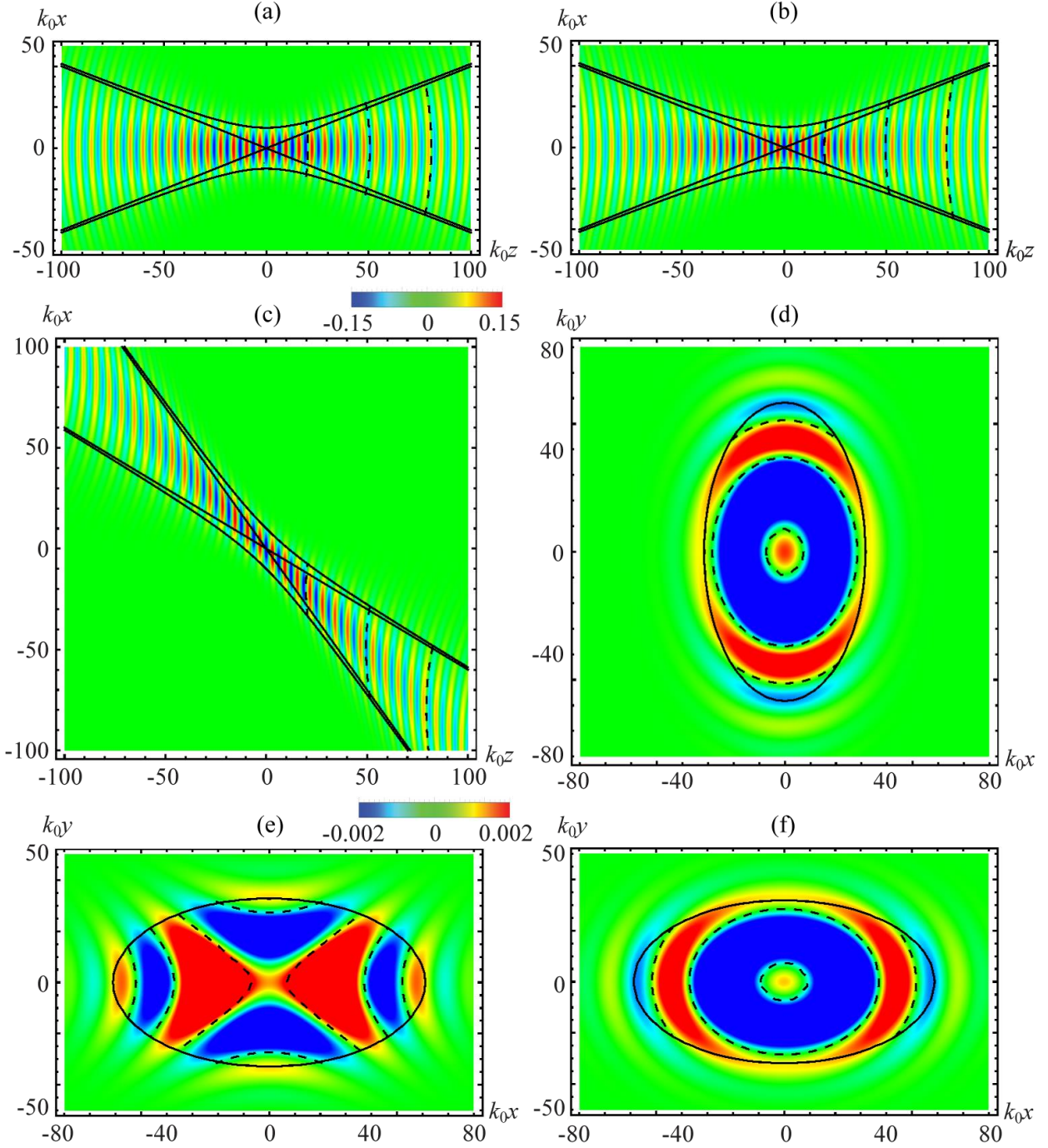

Fig. 1. (a)-(c) Plot of $\text{Re}\{g(x, t_x, r_x)e^{ik_z z}\}$ in $x$-$z$ plane color-coded above panel (c) for $k_z = 1$, $w_0 = 10$, and (a) $k_0 t_x = 0, k_0 r_x = -2$; (b) $k_0 t_x = 0, k_0 r_x = 2$; (c) $k_0 t_x = 1, k_0 r_x = 2$; (d)-(f) Plot of $\text{Re}\{g(x, t_x, r_x)g(y, t_x, r_x)e^{ik_z z}\}$ in $x$-$y$ plane color-coded above panel (e) for $k_0 t_x = k_0 t_y = 0, k_z/k_0 = 1, k_0 w_0 = 10$, and (d) $k_0 r_x = -1, k_0 r_y = -2, k_0 z = 100$; (e) $k_0 r_x = 2, k_0 r_y = -1, k_0 z = 105$; (f) $k_0 r_x = 2, k_0 r_y = 1, k_0 z = 101$; The constant phase curves Eq. (17) are dashed black and the constant field amplitude curves Eq. (18) are solid black.

The equation for constant phases is

$$\frac{k_z(x+t_x z)^2}{2R_x(z)} + \frac{k_z(y+t_y z)^2}{2R_y(z)} - \frac{1}{2}\psi_{G_x} - \frac{1}{2}\psi_{G_y} + k_z z = C_{phase} \tag{17}$$

Note that since, generally speaking, $r_x$ and $r_y$ are different at Fresnel wave surfaces in isotropy-broken media, the Rayleigh lengths $z_x$ and $z_y$, beam radii $w_x(z)$ and $w_y(z)$, and wavefront curvatures $R_x(z)$ and $R_y(z)$ are different as well. Correspondingly, the resulting Gaussian beams feature both intensity and phase astigmatism. This astigmatism is illustrated in Fig. 1 (d)-(e) for different signs of $r_x$ and $r_y$. While in all cases difference in magnitude of $r_x$ and $r_y$ leads to elliptical intensity spot, constant phase curves become hyperbolic if $r_x r_y < 0$.

From Eqs. (10)-(17) we see that wavefronts propagate at an angle to the beam, the phenomenon recently described by Durach as ray-wave tilt in isotropy-broken media in Ref. [1]. Ray-wave tilt can be seen as differential aberration of the ray and wave sources when seen in the material rest frame [1]. This is illustrated in Fig. 1(c), where wavefronts propagate at 45-degree angle to the beam, which possesses different ray and wave axes and represents a *biaxial Gaussian beam*.

In accordance with Ref. [1] wavefronts propagate in the direction of time averaged Minkowski momentum, whose density is given by $\boldsymbol{g}_{Min} = \frac{1}{4\pi c}\text{Re}\{\boldsymbol{D}^* \times \boldsymbol{B}\}$, i.e. along z-axis in notations above, while the beam propagates in the direction of the Abraham momentum whose time averaged density is $\boldsymbol{g}_{Abr} = \frac{1}{4\pi c}\text{Re}\{\boldsymbol{E}^* \times \boldsymbol{H}\}$. The surfaces of the constant field strength form hyperboloids along the axis $x = -t_x z$, $y = -t_y z$, i.e. in the direction of the Abraham momentum

$$\frac{(x+t_x z)^2}{w_0^2\left\{1+\left(\frac{z}{z_x}\right)^2\right\}} + \frac{(y+t_y z)^2}{w_0^2\left\{1+\left(\frac{z}{z_y}\right)^2\right\}} = C_{fstr} \tag{18}$$

The ray-wave tilt corresponds to the components of the hidden momentum $\Delta \boldsymbol{g} = \boldsymbol{g}_{Abr} - \boldsymbol{g}_{Min}$, which is transverse to the wavefront propagation direction [..]

$$\overline{\Delta \boldsymbol{g}_\perp} = -\frac{k_0}{ck^2}\left(\text{Re}\{i\rho_{be}\boldsymbol{D}^* + i\rho_{bm}\boldsymbol{B}^*\}\right) \tag{18}$$

According to Eq. (18) the ray-wave tilt is related to the propagation of electric and magnetic bound charge waves which are carried by the beam in the isotropy-broken media [1]. To illustrate this, we find the bound charges by investigating the longitudinal fields using Eq. (4)

$$(E_z, H_z)^T = \{-\widehat{M}_z^{-1} \cdot \widehat{M}_{z\|} + i\widehat{M}_z^{-1} \cdot (k_0^{-1}\widehat{\partial})\} \cdot (E_x, E_y, H_x, H_y)^T,$$

$$\widehat{\partial} = \begin{pmatrix} 0 & 0 & -\partial_{\rho y} & \partial_{\rho x} \\ \partial_{\rho y} & -\partial_{\rho x} & 0 & 0 \end{pmatrix}$$

$$\widehat{\partial}\begin{pmatrix} \boldsymbol{E}_\perp \\ \boldsymbol{H}_\perp \end{pmatrix} = \begin{pmatrix} 0 & 0 & -\partial_{\rho y} & \partial_{\rho x} \\ \partial_{\rho y} & -\partial_{\rho x} & 0 & 0 \end{pmatrix} \cdot (E_x, E_y, H_x, H_y)^T$$

Using the fact that the transverse gradient of $G(r)$ satisfies

$$\nabla_\rho G(r) = \frac{1}{2}\hat{\alpha}^{-1}\rho\, G(r) = -\frac{2}{w_0^2}\left(\frac{(x+t_x z)}{1+i\frac{z}{z_x}}, \frac{(y+t_y z)}{1+i\frac{z}{z_y}}\right) G(r)$$

we rewrite

$$\hat{\partial}\begin{pmatrix} E_\perp \\ H_\perp \end{pmatrix} = -\frac{2}{w_0^2}\begin{pmatrix} 0 & 0 & -\frac{(y+t_y z)}{1+i\frac{z}{z_y}} & \frac{(x+t_x z)}{1+i\frac{z}{z_x}} \\ \frac{(y+t_y z)}{1+i\frac{z}{z_y}} & -\frac{(x+t_x z)}{1+i\frac{z}{z_x}} & 0 & 0 \end{pmatrix}\begin{pmatrix} E_\perp \\ H_\perp \end{pmatrix} = \hat{P}(\rho)\begin{pmatrix} E_\perp \\ H_\perp \end{pmatrix}$$

Following this, the full field vectors of the biaxial Gaussian beam are

$$\begin{pmatrix} E \\ H \end{pmatrix} = G(r)\exp(ik_z z)\left\{\begin{pmatrix} E_0 \\ H_0 \end{pmatrix} + \hat{z}\,\widehat{M}_z^{-1}\cdot(ik_0^{-1}\hat{P})\begin{pmatrix} E_{\perp 0} \\ H_{\perp 0} \end{pmatrix}\right\}$$

$$= G(r)\exp(ik_z z)\left\{\begin{pmatrix} E_0 \\ H_0 \end{pmatrix} + \hat{z}\begin{pmatrix} \delta E_z(r) \\ \delta H_z(r) \end{pmatrix}\right\} \tag{18}$$

The bound charges can be found as

$$4\pi\rho_e = -4\pi\nabla\cdot P = \nabla\cdot E = \nabla\{G(r)\exp(ik_z z)\}\cdot E_0 + \partial_z\{G(r)\exp(ik_z z)\,\delta E_z(r)\}$$

$$4\pi\rho_m = -4\pi\nabla\cdot M = \nabla\cdot H = \nabla\{G(r)\exp(ik_z z)\}\cdot H_0 + \partial_z\{G(r)\exp(ik_z z)\,\delta H_z(r)\}$$

In Fig. 2 we consider a material with Fresnel waves surface $\mathcal{H}=0$ shown in Fig. 2(a) and the material parameters matrix $\widehat{M}$ in Fig. 2(b). It is a tetra-hyperbolic material. We consider a central plane wave with $n=1.915$ propagating along the z-axis, i.e. $g_{Min}\propto\hat{z}$ shown in Fig. 2(a) as a black dashed vector. The direction of the Abraham momentum and the normal to the Fresnel wave surface corresponds to $k_0 t_x = 2.21$ at the central wave vector. The curvature of the Fresnel wave surface is $k_0 r_x = 12.16$.

In Fig. 2(c) we plot $\mathrm{Re}\{g(x,t_x,r_x)e^{ik_z z}\}$ for a beam of width $k_0 w_0 = 20$ centered on this wave. As can be seen in Fig. 2(c) the ray is directed along the Abraham momentum, while the phase propagates along the Minkowski momentum. The wavefronts are converging after passing the focus of the beam, which corresponds to the positive curvature of the Fresnel wave surface at the central wave vector, which can be seen in Fig. 2(a).

In Fig. 2(d) and (e) we show the electric and magnetic bound charge waves which propagate along with the fields of the beam. Interestingly, while the bound magnetic charge has a diminished magnitude at the center of the beam, the bound electric charge is shifted towards the x-axis with respect to the fields.

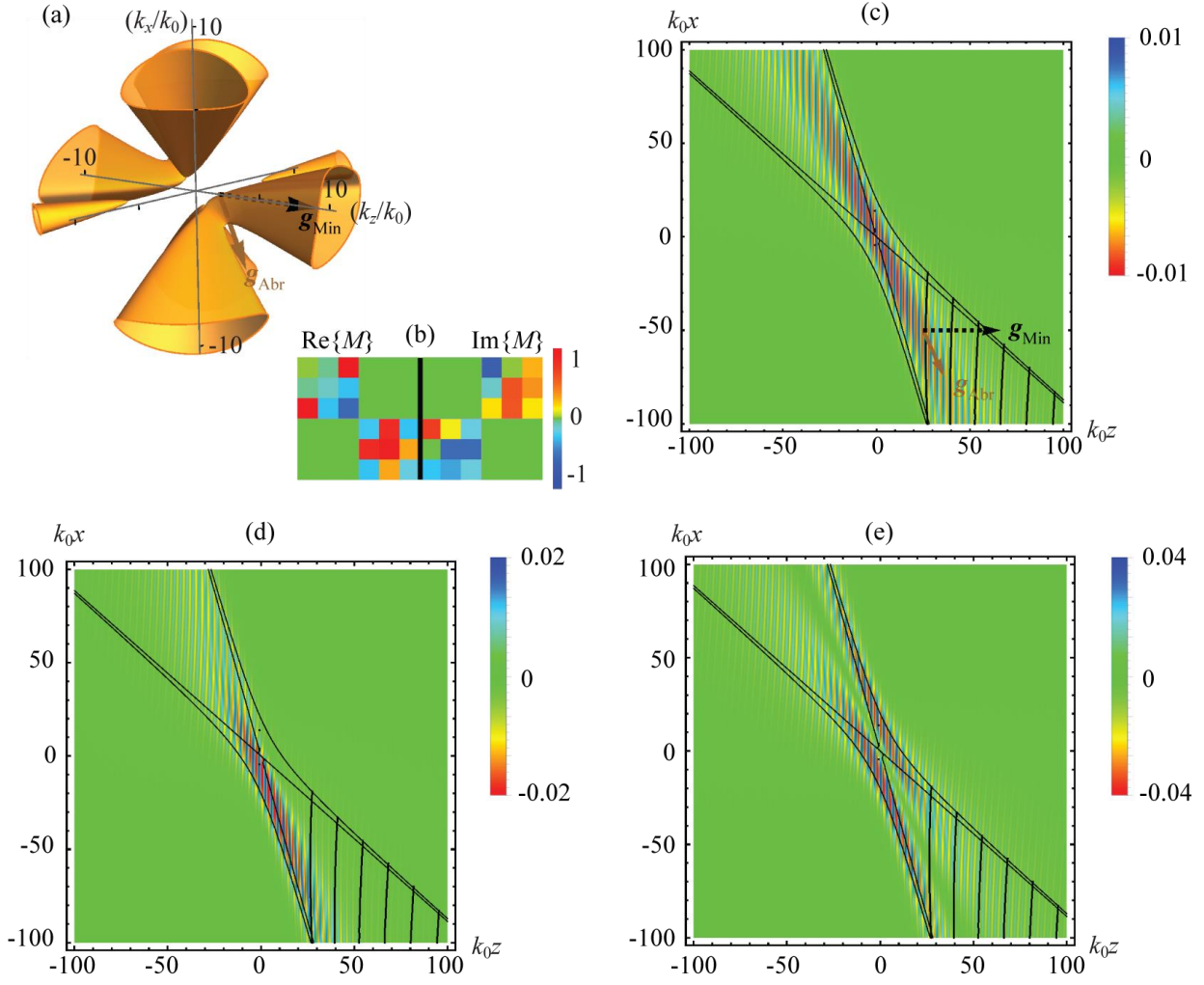

Fig. 2. (a) Isotropy-broken Fresnel wave surface for $\widehat{M}$ shown in next panel. Black dot indicates the central wave vector of the biaxial Gaussian beam and the corresponding Minkowski momentum (black dashed) and Abraham momentum (brown) indicate the directions of the axes; (b) isotropy-broken material parameters matrix $\widehat{M}$; (c) $\text{Re}\{g(x, t_x, r_x)e^{ik_z z}\}$ for the central wave vector shown in (a) with $k_0 w_0 = 20, k_0 t_x = 2.21, k_0 r_x = 12.16$; (d) $\text{Re}\{\rho_e\}$ the electric bound charge wave; (e) $\text{Re}\{\rho_m\}$ the magnetic bound charge wave.

### 3. Higher-order Biaxial Hermite-Gaussian Beams in Isotropy-Broken Media

To obtain the higher-order biaxial Hermite-Gaussian modes in the isotropy-broken media from the fundamental mode in Eqs. (14)-(16) we utilize the following operator

$$\left(\frac{\partial}{\partial \rho_x} + \alpha \rho_x\right)^n = \exp\left(-\frac{\alpha \rho_x^2}{2}\right)\left(\frac{\partial}{\partial \rho_x}\right)^n \exp\left(\frac{\alpha \rho_x^2}{2}\right), \quad \alpha = -\frac{2\, e^{i\psi_{G_x}}}{w_0\, w_x(z)}$$

The resulting higher order mode after excluding constants is expressed as

$$g_n(\rho_x) = g(\rho_x)\, e^{-in\psi_{G_x}}\, H_n\left(\frac{\sqrt{2}\rho_x}{w_x(z)}\right), \text{ or}$$

$$g_n(\rho_x) = \frac{1}{\pi^{1/2}}\frac{1}{\sqrt{w_x(z)}}\exp\left(-\frac{(x+t_x z)^2}{w_x^2(z)}\right) H_n\left(\frac{\sqrt{2}\rho_x}{w_x(z)}\right)\exp\left(i\left\{\frac{k_z(x+t_x z)^2}{2R_x(z)} - N\psi_{G_x}\right\}\right)$$

where the Hermite polynomials are $H_n(\xi) = (-1)^n \exp(\xi^2)\left(\frac{d}{d\xi}\right)^n \exp(-\xi^2)$ and $N = \left(n+\frac{1}{2}\right)$.

In Fig. 3 we plot the biaxial Hermite-Gaussian beams for the same parameters as in Fig. 1(c) for $n=1$ and $n=2$. We observe the transverse pattern of the field is perpendicular to the wavefront propagation direction, while the wavefronts are converging after passing the focal plane due to positive curvature of the Fresnel wave surface.

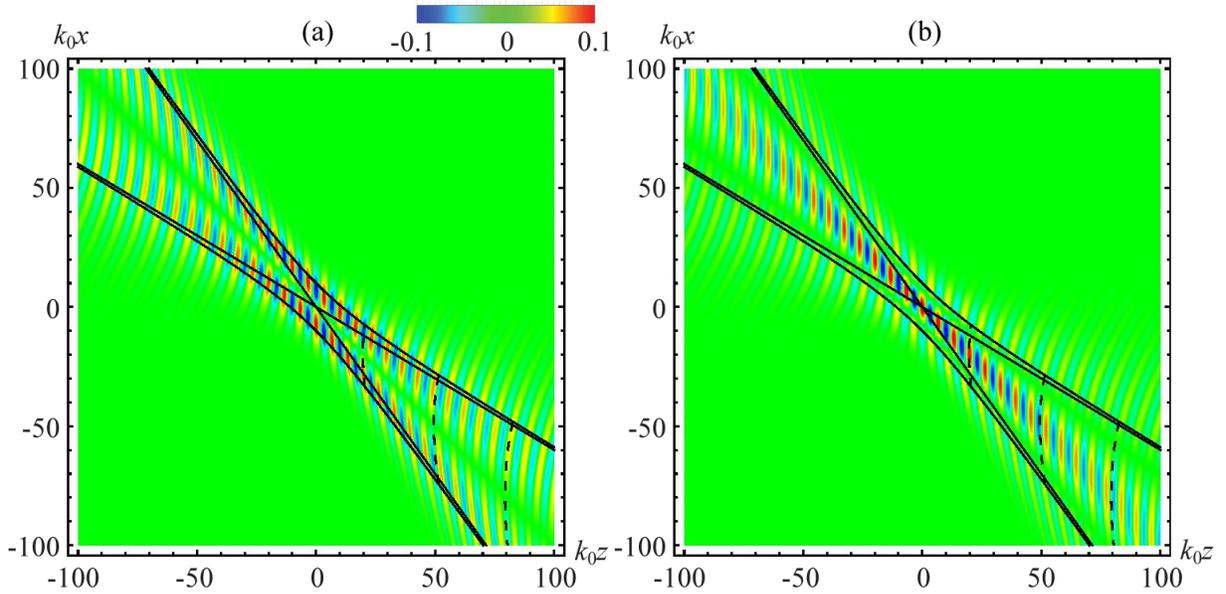

Fig. 3. Plot of the biaxial Hermite-Gaussian beams $\text{Re}\{g_n(x, t_x, r_x)e^{ik_z z}\}$ in x-z plane color-coded above the panels for the same parameters as in Fig. 1(c); (a) $n=1$; (b) $n=2$.

## 4. Laguerre-Gaussian Vortex Beams in Isotropy-Broken Media

Vortex beams has attracted a lot of attention in recent years due to their ability to carry and transfer OAM to materials. Consider a fundamental astigmatic biaxial Gaussian beam with equal Rayleigh lengths $z_x = z_y = z_R$, such that $w_x^2/w_y^2 = r_x/r_y$,

$$G(\mathbf{r}) = \frac{1}{(z - iz_R)}\exp\left(-i\frac{\left(\frac{\rho_x^2}{r_x} + \frac{\rho_y^2}{r_y}\right)}{2(z - iz_R)}\right)$$

From the fundamental mode we obtain higher order biaxial Laguerre-Gaussian vortex beams with radial indices $p \geq 0$ and azimuthal indices $l$

$$G_{pl}(\boldsymbol{r}) = \left(\frac{z+iz_R}{z-iz_R}\right)^p \left(\frac{1}{z-iz_R}\right)^{|l|+1} \left(\frac{\rho_x}{\sqrt{r_x}} - \frac{i\rho_y}{\sqrt{r_y}}\right)^{|l|} \exp\left(-i\frac{\left(\frac{\rho_x^2}{r_x}+\frac{\rho_y^2}{r_y}\right)}{2(z-iz_R)}\right) L_p^{|l|}\left[-z_R \frac{\left(\frac{\rho_x^2}{r_x}+\frac{\rho_y^2}{r_y}\right)}{z^2+z_R^2}\right]$$

In Fig. 4 we plot biaxial Laguerre-Gaussian vortex beams in isotropy-broken medium. We see that optical vortices are wrapped around the ray axis such that the transverse pattern is perpendicular to the wavefront propagation direction. The wavefronts are converging after passing the focal plane due to the positive curvature of the Fresnel wave surface at the central wave vector.

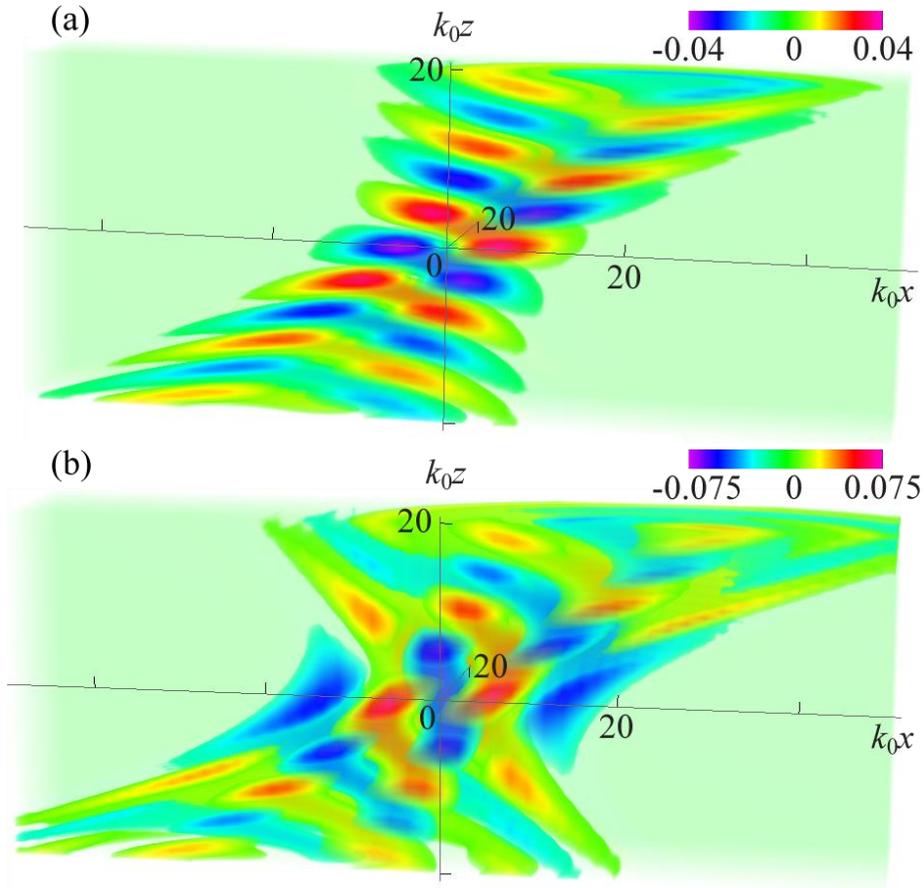

Fig. 4. Laguerre-Gaussian vortex beams Re$\{G_{pl}(\boldsymbol{r})e^{ik_z z}\}$ in isotropy-broken medium for $k_0 t_x = -1, k_0 z_R = -10, k_0 r_x = 3, k_0 r_y = 1$ with (a) $p = 0, l = 1$; (b) $p = 1, l = 2$.

Introducing the following parameters

$$\rho_L = \sqrt{-z_R\left(\frac{\rho_x^2}{r_x} + \frac{\rho_y^2}{r_y}\right)}, \varphi_L = \arctan\left(\frac{\rho_y\sqrt{r_x}}{\rho_x\sqrt{r_y}}\right),$$

$$w_L(z) = \sqrt{2}\sqrt{z^2 + z_R^2}, \psi_L = \arctan\left(\frac{z}{z_R}\right), R_L(z) = z\left(1 + \left(\frac{z_R}{z}\right)^2\right), \qquad N = 2p + |l|$$

and using a normalizing constant $C_{pl}$ we express the biaxial Laguerre-Gaussian vortex beams as

$$G_{pl} = \frac{C_{pl}}{w_L(z)}\left(\frac{\rho_L\sqrt{2}}{w_L(z)}\right)^{|l|+1} L_p^{|l|}\left[\frac{2\rho_L^2}{w_L^2(z)}\right]\exp\left(-\frac{\rho_L^2}{w_L(z)}\right)\exp\left(\frac{i\rho_L^2}{2z_R R_L(z)} + i(N+1)\psi_L\right)e^{-i|l|\phi}$$

5. *Conclusions*

We considered electromagnetic fields in isotropy-broken media in paraxial approximation. We derived the paraxial equation and obtained three classes of its solutions, namely, biaxial Gaussian, Hermite-Gaussian, and Laguerre-Gaussian beams. The parameters of the beams were directly related to the properties of the Fresnel wave surface at the central wave vector of the beam.

References.